# The equation of a Light Leptonic Magnetic Monopole and its Experimental Aspects

**Georges Lochak**
*Fondation Louis de Broglie*
*23, rue Marsoulan F-75012 Paris*
*inst.louisdebroglie@free.fr*

**Abstract**.
The present theory is closely related to Dirac's equation of the electron, but not to his magnetic monopole theory, except for his relation between electric and magnetic charge. The theory is based on the fact, that the *massless* Dirac equation admits a *second electromagnetic coupling*, deduced from a *pseudo-scalar* gauge invariance. The equation thus obtained has the symmetry laws of a massless *leptonic*, *magnetic monopole*, able to *interact weakly*. We give a more precise form of the Dirac relation between electric and magnetic charges and a quantum form of the Poincaré first integral. In the Weyl representation our equation splits into P-conjugated monopole and antimonopole equations with the correct electromagnetic coupling and *opposite chiralities*, predicted by P. Curie. Charge conjugated monopoles are *symmetric in space* and not in time (contrary to the electric particles) : an important fact for the vacuum polarization. Our monopole is a *magnetically excited neutrino*, which leads to experimental consequences. These monopoles are assumed to be produced by electromagnetic pulses or arcs, leading to nuclear transmutations and, for beta radioactive elements, a shortening of the life time and the emission of monopoles instead of neutrinos in a magnetic field. A corresponding discussion is given in section 15.

**1. Introduction.**

The hypothesis of separated magnetic poles is very old. In the 2$^{nd}$ volume of his famous *Treatise of Electricity and Magnetism* [1], devoted to Magnetism, Maxwell considered the existence of free magnetic charges as an evidence, just as the evidence of electric charges. He based the theory of magnetism on this hypothesis, and he reported that, as far back as 1785, Coulomb gave the *experimental proof* that the law of force of a magnetic charge is the same as the one of an electric charge : the well known *Coulomb law* . In his experiments, Coulomb took for a magnetic charge, the extremity of a thin magnetic rod. We quote only some papers on history : [2], [3], [4], later on, we shall restrict ourselves only to papers useful for our purpose. In the following we remain in the framework of electrodynamics, without including other monopoles such as the one of Dirac (which is independent of the equation of the electron) or the one of t'Hooft and Polyakov.

Contrary to the tendency to assume that a monopole must be heavy, bosonic, with strong interactions, without any symmetry law, our monopole appears as **a second application of the Dirac theory of the electron,** based on a **pseudo scalar gauge condition** from which we deduce **symmetry laws** predicted by Pierre Curie. Contrary to other theories, our monopole is light, fermionic and interacting electromanetically and weakly. It may be considered as a **magnetically excited neutrino.**

**2. The classical form of electromagnetic symmetries. The origin of the monopole.**

In his paper, *Symmetry in Physical Phenomena* [5], Pierre Curie put forward the constructive role of symmetry in physics. Generalizing the cristallographic groups, he defined the *invariance groups* of limited objects in R$^3$, and applied them to electromagnetism, only starting from *experiment* and not from the formal symmetry of the equations of electromagnetism. As a consequence of his laws, he infered the *possibility of* "*free magnetic charges*"[1][6].

The different symmetries of electric and magnetic charges are due to the fact that the electric field is a polar vector and the magnetic field is axial, which is proved experimentally [5]. For charges corresponding results have been proved the in the same way [7].

---

[1] In the reference [2], it is said that Curie *« suggests out of the blue »* that magnetic charge might exist. Probably, the authors have never seen the original Curie papers. Actually his prediction was a logical consequence of the symmetry laws of electromagnetism that he himself had discovered. It may be added that he made such a prediction for the second time : the first one was the theoretical prediction of *piezoelectricity* later observed by P. Curie. Such predictions were just as *« out of the blue »,* as the prediction of the neutrino by Pauli or of the antimatter by Dirac !

The scheme of *classical symmetries* for electromagnetic quantities is the following :

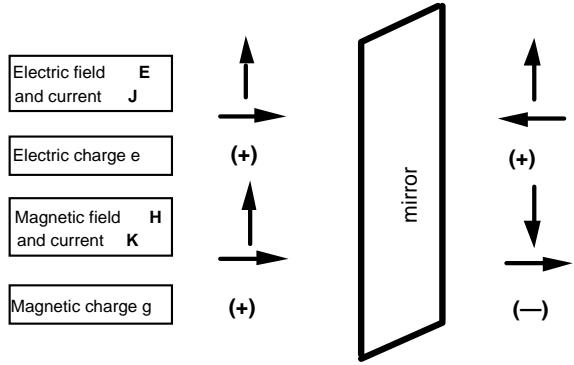

Fig. 1 Symmetry laws of electric and magnetic quantities

These symmetries are in accordance with those of **F**, **v** and of the laws of forces :

$$\mathbf{F}_{elec} = e\left(\mathbf{E} + 1/c\,\mathbf{v}\times\mathbf{H}\right); \quad \mathbf{F}_{magn} = g\left(\mathbf{H} - 1/c\,\mathbf{v}\times\mathbf{E}\right) \qquad (2.1)$$

and they entail polar and axial transformations of electric and magnetic currents :

$$\mathbf{J} = e\,\mathbf{v}, \quad \mathbf{K} = g\,\mathbf{v} \qquad (2.2)$$

*Nevertheless, it is shocking that in virtue of Curie laws, the magnetic charge g is a pseudoscalar,* because a physical constant has no tensorial transformations (*c* does not vary as a velocity and the Planck constant *h* does not vary as a kinetic moment). It will be shown in quantum mechanics, that *the magnetic charge is a scalar* ($P : g \to g$) while the *pseudo-scalar* transformations is not the property of the charge, but of a *charge operator*.

The magnetic current will be an axial vector, like in (2.2), but with another definition. The Fig.1 is true, except for the magnetic charge. This is important because, according to a classical objection, magnetic poles could be eliminated from Maxwell's equations by a linear transformation. Denoting the fields as $(\mathbf{E}, \mathbf{H})$, the electric and magnetic currents as $(\mathbf{J}, \mathbf{K})$ and the electric and magnetic densities as $(\rho, \mu)$, the invariant linear transformation is :

$$\begin{aligned}
\mathbf{E} &= \mathbf{E}'\cos\gamma + \mathbf{H}'\sin\gamma\,; \quad \mathbf{H} = -\mathbf{E}'\sin\gamma + \mathbf{H}'\cos\gamma \\
\rho &= \rho'\cos\gamma + \mu'\sin\gamma\,; \quad \mu = -\rho'\sin\gamma + \mu'\cos\gamma \\
\mathbf{J} &= \mathbf{J}'\cos\gamma + \mathbf{K}'\sin\gamma\,; \quad \mathbf{K} = -\mathbf{J}'\sin\gamma + \mathbf{K}'\cos\gamma
\end{aligned} \qquad (2.3)$$

By a choice of $\gamma$, **K** could be so eliminated from the equations, but *only* if **J**′ and **K**′ were *colinear*, and we shall see, that it cannot happen in our theory [8].

**3. The Birkeland-Poincaré effect.**

In 1896, Birkeland introduced a magnet in a Crookes' tube and he found a focusing of the cathodic beam [9]. Poincaré ascribed this effect to the force of a magnetic pole at rest on a moving electric charge[8], [10] and he found the equation :

$$\frac{d^2\mathbf{r}}{dt^2} = \lambda\,\frac{1}{r^3}\,\frac{d\mathbf{r}}{dt}\times\mathbf{r}\,; \quad \lambda = \frac{e\,g}{mc} \qquad (3.1)$$

where *e* and *m* are the electric charge and the mass of the cathodic particles (*electrons*).

Poincaré showed that **r** follows a *geodesic line* of an axially symmetric cone (the *Poincaré cone*) and he proved the observed *focusing* effect. This is an important result because Coulomb proved that his law is the same for electricity and magnetism.

Therefore, a classical magnetic monopole in a Coulomb electric field obeys the Poincaré equation. Later on we shall find that *this equation is the classical limit of our equation* [8], [12]-[14]. Therefore, the fact that the Birkeland effect is predicted by the equation (3.1), becomes an argument in favor of our quantum equation.

Let us add two remarks :

1) Poincaré deduced from his equation, the angular momentum $\mathbf{J} = m\Lambda$ :

$$\Lambda = \mathbf{r} \times \frac{d\mathbf{r}}{dt} + \lambda \frac{\mathbf{r}}{r} \tag{3.2}$$

J.J. Thomson showed that the second term is the electromagnetic momentum [8], [11].

2) The Poincaré cone is the envelope of the *symmetry axis* $\mathbf{r}$ (joining electric and magnetic charges), rotating under a constant angle $\Theta'$, around the *momentum* $\mathbf{J} = m\Lambda$. But this is the definition of the *Poinsot cone* of a symmetric top. Thus, we can deduce that *the system of an electric and a magnetic charge has the symmetry of a symmetric top [8], [15]*. This will be important later.

Owing the following properties all that was said is summarized on the Fig. 2.

$$\frac{d^2\mathbf{r}}{dt^2} \cdot \mathbf{r} = \frac{d^2\mathbf{r}}{dt^2} \cdot \frac{d\mathbf{r}}{dt} = 0 \; ; \; \Lambda \mathbf{r} = \lambda r \tag{3.3}$$

Our equation of a monopole will define this cone in a quantum form.

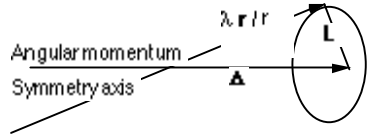

Fig. 2 The generation of the Poincaré (or Poinsot) cone and the decomposition of the total momentum.

**4. The electromagnetic potentials for a magnetic pole.**

Let us write the Maxwell equations with the electric and magnetic currents $(\mathbf{J}, \mathbf{K})$ and charges $(\rho, \mu)$:

$$curl\ \mathbf{H} - \frac{1}{c}\frac{\partial \mathbf{E}}{\partial t} = \frac{4\pi}{c}\mathbf{J} \; ; \; -curl\ \mathbf{E} - \frac{1}{c}\frac{\partial \mathbf{H}}{\partial t} = \frac{4\pi}{c}\mathbf{K}$$

$$div\ \mathbf{E} = 4\pi\rho ; \; div\ \mathbf{H} = 4\pi\mu \tag{4.1}$$

In relativistic coordinates :

$$x^\alpha = \{x^1, x^2, x^3, x^4\} = \{x, y, z, ict\} \tag{4.2}$$

the equations (4.1) become :

$$\partial^\beta F_{\alpha\beta} = \frac{4\pi}{c}J_\alpha \; ; \; J_\alpha = (\mathbf{J}, i\rho c) \; ; \; \partial^\beta \overline{F}_{\alpha\beta} = \frac{4\pi}{c}K_\alpha \; ; \; iK_\alpha = (\mathbf{K}, i\mu c) \tag{4.3}$$

where : $\overline{F}_{\alpha\beta} = \frac{i}{2}\varepsilon_{\alpha\beta\gamma\delta}F^{\gamma\delta}$ $(\varepsilon_{\alpha\beta\gamma\delta}\ antisymmetric)$.

It is clear that we cannot define the field by a Lorentz *polar* potential, because :

$$F_{\alpha\beta} = \partial_\alpha A_\beta - \partial_\beta A_\alpha \Rightarrow \frac{1}{2}\varepsilon_{\alpha\beta\gamma\delta}\partial^\beta F^{\gamma\delta} = 0 \tag{4,4}$$

Therefore, we must introduce a new potential $B_\alpha$ such that :

$$F_{\alpha\beta} = \frac{1}{2} \varepsilon_{\alpha\beta\gamma\delta} \left( \partial_\gamma B_\delta - \partial_\delta B_\gamma \right) \tag{4.5}$$

$B_\alpha$ must be a *pseudo-potential*, the dual of an antisymmetric tensor of the third order :

$$B_\alpha = \frac{1}{3!} \varepsilon_{\alpha\beta\gamma\delta} C^{\beta\gamma\delta} \tag{4.6}$$

In terms of ordinary coordinates, we have :

$$A_\alpha = (\mathbf{A}, iV); \; iB_\alpha = (\mathbf{B}, iW) \tag{4.7}$$

where **B** is an axial vector and $W$ a pseudoscalar. The fields are defined as :

$$\mathbf{E} = curl\,\mathbf{B}; \; \mathbf{H} = \nabla W + \frac{1}{c}\frac{\partial \mathbf{B}}{\partial t} \tag{4.8}$$

The preceeding formulae were at first given by de Broglie [16] and later related to the monopole by Cabibbo and Ferrari [17].

**5. Dirac strings :**

In 1931, Dirac raised the problem of the motion of an electric charge around a fixed monopole or conversely [18]. In the case of the motion of a monopole in the vicinity of an electric coulombian center, the electric field **E** of the latter will be defined by a pseudo potential **B** deduced from (4.8) :

$$curl\,\mathbf{B} = e\,\frac{\mathbf{r}}{r^3} \tag{5.1}$$

**B** cannot be continuous and uniform. There must be a singular line : the *Dirac string* and to save the uniformity of wave functions, Dirac found his famous relation between electric and magnetic elementary charges (see [8], [18] for the Dirac proof) :

$$D = \frac{eg}{\hbar c} = \frac{n}{2} \tag{5.2}$$

Later on we shall give a proof based on our equation [8]. Let us note two points :
- In the Dirac proof, the string plays the central role. On the contrary, in our proof the string will be rubbed out by an argument of symmetry.
- Dirac's choice of potentials corresponds to the following solution of (5,1) which has no defined symmetry and makes the calculations more difficult.

$$B_x = \frac{e}{r}\frac{-y}{r+z},\; B_y = \frac{e}{r}\frac{x}{r+z},\; B_z = 0 \; \left( r = \sqrt{x^2+y^2+z^2} \right) \tag{5.3}$$

In the following, we shall chose another gauge that gives a pseudo-vectorial potential in accordance with the symmetry of the problem, which allows simplified calculations. This potejntial is :

$$B_x = \frac{e}{r}\frac{yz}{x^2+y^2},\; B_y = \frac{e}{r}\frac{-xz}{x^2+y^2},\; B_z = 0 \; \left( r = \sqrt{x^2+y^2+z^2} \right) \tag{5.4}$$

**6. Symmetry in quantum form.**

The main problem of the magnetic monopole was discovered by Maxwell [1] and Pierre Curie [5], [6] : it is the difference of symmetry between electricity and magnetism, i.e. between polar and axial vectors. This is the starting point of the following theory, which is based on the fact that Dirac's equation of the electron admits not only one *local gauge* but two, and only two.

The first invariance corresponds to an electric charge, the second one to a magnetic monopole. The new spinorial equations so obtained describe the Curie symmetry laws, in quantum terms. These laws indeed clearly appear *only in quantum mechanics*.

**a) The two gauges of Dirac's equation.**

Let us write the Dirac equation without external field :

$$\gamma_\mu \partial_\mu \Psi + \frac{m_0 c}{h} \Psi = 0 \qquad (6.1)$$

We shall use the de Broglie representation which giçves a plus sign in $\gamma_5$ :

$x_\mu = \{x_k \; ; \; i\,c\,t\}$, $\gamma_\mu$ are defined in terms of Pauli matrices $s_k$ as :

$$\gamma_k = i \begin{pmatrix} 0 & s_k \\ -s_k & 0 \end{pmatrix} ; \; k = 1, 2, 3 ; \; \gamma_4 = \begin{pmatrix} I & O \\ O & -I \end{pmatrix} ; \; \gamma_5 = \gamma_1 \gamma_2 \gamma_3 \gamma_4 = \begin{pmatrix} 0 & I \\ I & 0 \end{pmatrix} \qquad (6.2)$$

Consider the following global gauge, where θ is a constant phase and Γ a hermitian matrix that will be represented in Clifford algebra basis :

$$\Psi \to e^{i\Gamma\theta}\Psi \quad (\Gamma = \sum_{N=1}^{16} a_N \Gamma_N ; \; \Gamma_N = \{I, \gamma_\mu, \gamma_{[\mu}\gamma_{\nu]}, \gamma_{[\lambda}\gamma_\mu\gamma_{\nu]}, \gamma_5\}) \qquad (6.3)$$

Introducing this gauge in (6.1), we get :

$$\left(\gamma_\nu e^{i\Gamma\theta}\gamma_\nu\right)\gamma_\mu \partial_\mu \Psi + \frac{m_0 c}{h} e^{i\Gamma\theta}\Psi = 0 \qquad (6.4)$$

Developing Γ as in (6.3) and using the equality $\gamma_\mu \Gamma_N \gamma_\mu = \pm\Gamma_N$ [19], we find :

$$\gamma_\mu e^{i\Gamma\theta}\gamma_\mu = \exp\left(i\,\theta \sum_{N=1}^{16} \pm a_N \gamma_\mu \Gamma_N \gamma_\mu\right) = \exp\left(i\,\theta \sum_{N=1}^{16} \pm a_N \Gamma_N\right) \qquad (6.5)$$

(6.1) will be invariant if $\gamma_\mu e^{i\Gamma\theta}\gamma_\mu$ commutes or anticommutes with all the $\gamma_\mu$ i.e. if $\Gamma = \mathbf{I}$ or $\Gamma = \gamma_5$:

$$if \; \Gamma = \mathbf{I}, \; \Psi \to e^{i\theta}\Psi \; ; \; if \; \Gamma = \gamma_5, \; \Psi \to e^{i\gamma_5\theta}\Psi \qquad (6.6)$$

The first case is the *phase invariance* which gives the *conservation of electricity*. The second case will be called *chiral invariance* and will give the *conservation of magnetism*.

But the first one is valid for every value of $m_0$ in eq. (6.1), so that the conservation of electricity is universal in quantum mechanics ; the second one (which was given in [20], [21], [22]) is valid only for $m_0 = 0$ because of the anticommutation of $\gamma_5$ and $\gamma_\mu$, so that *the conservation of magnetism is weaker than conservation of electricity*.

**b) The Dirac tensors and the magic angle A of Yvon-Takabayasi.**

In the Clifford basis (6.3), the Dirac spinor defines 16 tensorial quantities. A scalar, a polar vector, an antisymmetric tensor of rank two, an antisymmetric tensor of rank three (an axial vector) and an antisymmetric tensor of rank four (a pseudo-scalar) :

$$\Omega_1 = \bar{\Psi}\Psi \,;\, J_\mu = i\,\bar{\Psi}\gamma_\mu\Psi \,;\, M_{\mu\nu} = i\,\bar{\Psi}\gamma_\mu\gamma_\nu\Psi \,;\, \Sigma_\mu = i\,\bar{\Psi}\gamma_\mu\gamma_5\Psi \,;\, \Omega_2 = -i\,\bar{\Psi}\gamma_5\Psi$$
$$\left(\bar{\Psi} = \Psi^+\gamma_4;\ \Psi^+ = \Psi\,h.c.\right) \tag{6.7}$$

If $\Omega_1$ and $\Omega_2$ do not vanish simultaneously, the Dirac spinor may be written as [12], [20], [21] :

$$\Psi = \rho\, e^{i\gamma_5 A}\, U\, \Psi_0 \tag{6.8}$$

$\rho$ = amplitude, $A$ = *pseudo-scalar angle* of Yvon-Takabayasi, $U$ = general Lorentz transformation, $\Psi_0$ = constant spinor, and :

$$\rho = \sqrt{\Omega_1^2 + \Omega_2^2}\ ;\ A = Arc\,tg\,\frac{\Omega_2}{\Omega_1} \tag{6,9}$$

$U$ is a product of six factors $e^{i\Gamma\vartheta}$ with three real Euler angles (rotations in $R^3$) and three imaginary angles (components of velocity). The proper rotation Euler angle $\varphi$ gives a *scalar phase* $\tilde{\varphi}/2$ in the spinor $\Psi$, conjugated (by a *classical Poisson bracket*) to the component $J_4$ of the *polar vector* $J_\mu$ ; the *pseudo-scalar* angle $A$ is conjugated to the component $\Sigma_4$ of the *axial vector* $\Sigma_\mu$ ( [20], [21], [22]). So we have the classical field poisson brackets [20] :

$$\left[\frac{\varphi}{2},\, J_4\right] = \delta(\mathbf{r}-\mathbf{r}')\ ;\ \left[\frac{A}{2},\, \Sigma_4\right] = \delta(\mathbf{r}-\mathbf{r}') \tag{6.10}$$

In the Dirac theory, $J_4$ is a density of *electricity* associated to the *phase invariance* ; the spatial part of $J_\mu$ is a density of electric current. $\Sigma_4$ is a density associated in the same way to the chiral invariance (6.6) and it will be shown that the space part of $\Sigma_\mu$ is a density of magnetic current. So, there are densities of *magnetic* charge and current. The difference between the two gauges is :
1) $J_\mu$ is *polar* and $\Sigma_\mu$ *axial*,
2) $J_\mu$ is *time-like* and $\Sigma_\mu$ is *space-like* because of the Darwin - de Broglie equalities :

$$-J_\mu J_\mu = \Sigma_\mu \Sigma_\mu = \Omega_1^2 + \Omega_2^2\ ;\ J_\mu \Sigma_\mu = 0 \tag{6.11}$$

It is because $J_\mu$ is *time-like*, that it may be a current of electricity and probability. Thus it seems that a space-like $\Sigma_\mu$ will be unacceptable. We shall see that this is not the case.

**c) PTC symmetries of the angle A.**

It was proved [7] that, the correct transformations, in the sense of Curie, are such that **P** is a Racah transformation, but **T** is not, it is the antilinear « weak time reversal »[2] :

$$\begin{aligned}\mathbf{P} &:\ \Psi \to \gamma_4 \Psi \\ \mathbf{T} &:\ \Psi \to -i\,\gamma_3\gamma_1\Psi^*,\ (e \to -e) \\ \mathbf{C} &:\ \Psi \to \gamma_2\Psi^*,\ (e \to -e)\end{aligned} \tag{6.12}$$

With the definitions (6.9), this implies :

---

[2] While the Racah transformation would be linear : $\psi \to \gamma_1\gamma_2\gamma_3\psi$.

$$\mathbf{P} : \Omega_1 \to \Omega_1 ; \Omega_2 \to -\Omega_2$$
$$\mathbf{T} : \Omega_1 \to \Omega_1 ; \Omega_2 \to -\Omega_2 \qquad (6.13)$$
$$\mathbf{C} : \Omega_1 \to -\Omega_1 ; \Omega_2 \to -\Omega_2$$

(6.9) and (6.11) show that $A$ is a relativistic pseudo-invariant which is **PTC** invariant. Owing to (6.9), we can give a geometrical interpretation of the chiral gauge, writing:

$$\Omega_1 = \rho \cos A ; \quad \Omega_2 = \rho \sin A \qquad (6.16)$$

Now, consider a chiral gauge, slightly modified with respect to (6,6):

$$\Psi' = e^{i \gamma_5 \theta / 2} \Psi \qquad (6.17)$$

Using the definition (6,11) of $\Omega_1$ and $\Omega_2$, we get from (6,16):

$$\begin{pmatrix} \Omega_1' \\ \Omega_2' \end{pmatrix} = \begin{pmatrix} \cos \theta & -\sin \theta \\ \sin \theta & \cos \theta \end{pmatrix} \begin{pmatrix} \Omega_1 \\ \Omega_2 \end{pmatrix} \qquad (6.18)$$

The chiral transformation is a rotation $\theta$ in the plane $\{\Omega_1, \Omega_2\}$, that will be called the *chiral plane*, or a rotation $\theta / 2$ of the spinor. And (6.16) shows that θ is a phase shift of the angle A:

$$A' = A + \theta \qquad (6.19)$$

**d) The wave equation.**

We know that the *local* gauge deduced from the global first gauge (6,8) gives the minimal electric coupling in the Dirac equation. Now, consider the Dirac equation with $m_0 = 0$:

$$\gamma_\mu \partial_\mu \Psi = 0 \qquad (6.20)$$

It is invariant by the chiral gauge (6.8). Let us introduce a pseudo-scalar phase $\phi$, the corresponding gauge transformation and the charge operator $G$:

$$\Psi \to \exp\left(i \frac{g}{\hbar c} \gamma_5 \phi\right) \Psi ; \quad B_\mu \to B_\mu + i \partial_\mu \phi ; \quad G = g \gamma_5 \qquad (6.21)$$

g is a *scalar* magnetic charge : *the pseudo-scalar character of magnetism is related to a pseudo-scalar magnetic charge operator $G$ which is at the origin of all the differences between the classical and the quantum theory of magnetic monopoles*.

ϕ being a pseudo-scalar, the potential is not a polar vector $R =$, but an *axial potential $B_\mu$* defined in (4.6), (4.7), which has the variance of $\partial_\mu \phi$. The covariant derivatives are:

$$\nabla_\mu = \partial_\mu - \frac{g}{\hbar c} \gamma_5 B_\mu \left(= \partial_\mu - \frac{G}{\hbar c} B_\mu\right) \qquad (6.22)$$

where the absence of $i$ in front of g is due to the axiality of $B_\mu$. **The equation of the magnetic monopole** is thus [12], [13], [14]:

$$\gamma_\mu \left(\partial_\mu - \frac{g}{\hbar c} \gamma_5 B_\mu\right) \Psi = 0 \qquad (6.23)$$

## 7. Symmetries of the wave equation.

### a) Gauge invariance.

(6.23) is gauge invariant by (6.21). This entails the conservation of the *axial current* that plays the role of a magnetic current :

$$\partial_\mu K_\mu = 0 \; ; \; K_\mu = g \Sigma_\mu = i g \, \overline{\Psi} \gamma_\mu \gamma_5 \Psi \tag{7.1}$$

According to (6,11), the magnetic current cannot be colinear to the electric current, which prevents the application of (2.3) to remove $\left(K_\mu, \rho\right)$ in (2.3)). $K_\mu$ is a pseudo-tensor, as it was predicted by Curie. The space-like character will become clear a little further. This expression for the magnetic current was suggested by Salam [23], for symmetry reasons but here, it is a *consequence* of a wave equation and a gauge condition.

### b) CPT.

It is easy to prove that the wave equation (6,23) is C, P and T invariant [7][3] :

$$P : g \to g \; ; \; x_k \to -x_k \; ; \; x_4 \to x_4 \; ; \; B_k \to B_k \; ; \; B_4 \to -B_4 \; ; \; \Psi \to \gamma_4 \Psi$$
$$T : g \to g \; ; \; x_k \to x_k \; ; \; x_4 \to -x_4 \; ; \; B_k \to -B_k \; ; \; B_4 \to B_4 \; ; \; \Psi \to -i \gamma_3 \gamma_1 \Psi^* \tag{7.2}$$
$$C : g \to g \; ; \; \Psi \to \gamma_2 \Psi^*$$

In these formulae, an important point is that *the charge conjugation does not change the sign of the magnetic constant of charge g*. In the next section, we shall see what means the charge conjugation in the magnetic case. We can already assert that two conjugated monopoles have the same charge constant. Two monopoles with opposite constants are not charge conjugated : to change g in – g in (6.23) means *change the vertex angle of the Poincaré cone*.

We cannot create or annihilate pairs of monopoles with charges *g and – g*, as it is the case for electric charges *e and – e*. This property of charge conjugation of (6.23), shows that there is no danger of an infinite polarization of vacuum with such zero mass monopoles. Moreover, it shows that *one has not to invoke the great masses* to explain the rarity of monopoles or the difficulty to observe them.

The fact that chiral invariance and conservation of magnetism are easily broken, suggests that, more probably, monopoles are abundant in nature and that the problem of the isolation of one of them is not a problem of energy.

## 8. Weyl's representation. Two-component theory.

The matrix $\gamma_5$ and the magnetic charge operator G are diagonalized in the Weyl representation, and the wave function is divided into the two-component spinors ξ and η. So we have :

$$\Psi \to U\Psi = \begin{pmatrix} \xi \\ \eta \end{pmatrix} \; ; \; U = U^{-1} = \frac{1}{\sqrt{2}} (\gamma_4 + \gamma_5) \tag{8.1}$$

$$UGU^{-1} = Ug\gamma_5 U^{-1} = g\gamma_4 = \begin{pmatrix} g & 0 \\ 0 & -g \end{pmatrix} \tag{8.2}$$

(8.2) and (8.1) show that ξ and η are eigenstates of G, with the eigenvalues $\tilde{g}$ and $-g$ :

---

[3] In [12], [13], [14]), I gave the Racah formula for T, but it contradicts the Curie laws [7]. So, I have adopted the law : $g \to g$, $\Psi \to -i \gamma_3 \gamma_1 \Psi^*$ in the magnetic case.

$$UGU^{-1} \begin{pmatrix} \xi \\ 0 \end{pmatrix} = g \begin{pmatrix} \xi \\ 0 \end{pmatrix} \;;\; UGU^{-1} \begin{pmatrix} 0 \\ \eta \end{pmatrix} = g \begin{pmatrix} 0 \\ -\eta \end{pmatrix} \tag{8.3}$$

Owing to (8.1) and (4.7), (6.23) *splits into a pair* of uncoupled two-component equations in $\xi$ and $\eta$, corresponding to *opposite eigenvalues* of G [8], [13], [14] :

$$\left[ \frac{1}{c} \frac{\partial}{\partial t} - \mathbf{s}.\nabla - i \frac{g}{\hbar c} (W + \mathbf{s}.\mathbf{B}) \right] \xi = 0$$
$$\left[ \frac{1}{c} \frac{\partial}{\partial t} + \mathbf{s}.\nabla + i \frac{g}{\hbar c} (W - \mathbf{s}.\mathbf{B}) \right] \eta = 0 \;,\; i B_\mu = (\mathbf{B}, i W) \tag{8.4}$$

**P** and **T** permute the equations (8,4) between themselves and **P**, **T**, **C** become :
$$P : g \to g \;;\; x_k \to - x_k \;;\; t \to t \;;\; B_k \to B_k \;;\; W \to -W; \xi \leftrightarrow \eta$$
$$T : g \to g \;;\; x_k \to x_k \;;\; t \to -t \;;\; B_k \to -B_k \;;\; W \to W; \xi \to s_2 \xi^*; \eta \to s_2 \eta^* \tag{8.5}$$
$$C : g \to g \;;\; \xi \to -i s_2 \eta^*; \eta \to i s_2 \xi^*$$

We have a pair of *charge conjugated* particles — a *monopole* and an *antimonopole* — with the same charge constant but *opposite helicities*. They are defined by the operator G, which shows that **our monopole is a magnetically *excited neutrino, because*** (8.4) reduces to a pair of two-component neutrino equations if g = 0 [8], [13], [14].

The eq. (8.4) are invariant under the following gauge transformation (with opposite signs of the phase $\phi$ for $\xi$ and $\eta$) :

$$\xi \to \exp\left( i \frac{g}{\hbar c} \phi \right) \xi; \eta \to \exp\left( -i \frac{g}{\hbar c} \phi \right) \eta; W \to W + \frac{1}{c} \frac{\partial \phi}{\partial t}; \mathbf{B} \to \mathbf{B} - \nabla \phi \tag{8.6}$$

**9. Chiral currents.**

The Gauge (8.6) entails for (8.4) the *conservation laws* :

$$\frac{1}{c} \frac{\partial (\xi^+ \xi)}{\partial t} - \nabla \xi^+ \mathbf{s}\, \xi = 0 \;;\; \frac{1}{c} \frac{\partial (\eta^+ \eta)}{\partial t} + \nabla \eta^+ \mathbf{s}\, \eta = 0 \tag{9.1}$$

We thus have two currents, with some simple but important properties :

$$X_\mu = (\xi^+ \xi, -\xi^+ \mathbf{s}\, \xi);\; Y_\mu = (\eta^+ \eta,\, \eta^+ \mathbf{s}\, \eta);\; X_\mu X_\mu = 0 \;;\; Y_\mu Y_\mu = 0 \;;\; \mathbf{P} \Rightarrow X_\mu \leftrightarrow Y_\mu \tag{9.2}$$

They are *isotropic* and they are interchanged by parity : they are *chiral currents*.

Owing to (8.1), we find a decomposition of the polar and axial vectors defined in (6.9) :

$$J_\mu = X_\mu + Y_\mu \;;\; \Sigma_\mu = X_\mu - Y_\mu \tag{9.5}$$

*The chiral currents $X_\mu$ and $Y_\mu$ , may be taken as fundamental currents, that define electric and magnetic currents*. And we can prove (6.13), using (6.11) and (8.1), which gives :

$$\Omega_1 = \xi^+ \eta + \eta^+ \xi;\; \Omega_2 = i (\xi^+ \eta - \eta^+ \xi);\; \rho^2 = 4 (\xi^+ \eta)(\eta^+ \xi) \tag{9.6}$$

The fact that one of the vectors $J_\mu$, $\Sigma_\mu$ is time-like and the other space-like is a trivial property of the addition of isotropic vectors. But the fact that, *precisely $J_\mu$* is space-like, is a specific property due to the value of $\Omega_1^2 + \Omega_2^2$. See (6.13) and (9.6).

*Our magnetic current* $\mathbf{K}_\mu = g \Sigma_\mu$, *may be space-like* because the true magnetic currents are the isotropic currents $g X_\mu$ and $\tilde{g} Y_\mu$, whereas $\mathbf{K}_\mu$ *is only their difference*, which has not any reason to be of a definite type.

**10. The geometrical optics approximation and the Poincaré equation [8]**

Now we verify that we can find the Poincaré equation and the Birkeland effect. Let us introduce in the first equation (8.4) the following expression of the spinor $\xi$ :

$$\xi = a\, e^{iS/\hbar} \tag{10.1}$$

where $a$ is a two-component spinor and $S$ a phase. At zero order in $\hbar$, we have :

$$\left[\frac{1}{c}\left(\frac{\partial}{\partial t} - gW\right) - \left(\nabla S + \frac{g}{c}\mathbf{B}\right)\cdot \mathbf{s}\right] a = 0 \tag{10.2}$$

A *condition* for a non trivial solution $a$ is a relativistic zero mass Jacobi equation :

$$\frac{1}{c^2}\left(\frac{\partial}{\partial t} - gW\right)^2 - \left(\nabla S + \frac{g}{c}\mathbf{B}\right)^2 = 0 \tag{10.3}$$

We can define the kinetic energy $E$, the impulse $\mathbf{p}$, the linear Lagrange momentum $\mathbf{P}$ and the hamiltonian function $H$ :

$$E = -\frac{\partial S}{\partial t} + gW;\ \mathbf{p} = \nabla S + \frac{g}{c}\mathbf{B};\ \mathbf{P} = \nabla S\ ;H = c\sqrt{\left(\mathbf{P} + \frac{g}{c}\mathbf{B}\right)^2} - gW \tag{10.4}$$

A classical calculation gives the equation of motion :

$$\frac{d\mathbf{p}}{dt} = g\left(\nabla W + \frac{\partial \mathbf{B}}{\partial t}\right) - \frac{g}{c}\mathbf{v} \times curl\ \mathbf{B} \tag{10.5}$$

and the formulae (4.8) give the classical form :

$$\frac{d\mathbf{p}}{dt} = g\left(\mathbf{H} - \frac{1}{c}\mathbf{v} \times E\right) \tag{10.6}$$

Now, one should remember that the mass of the particle is equal to zero, so that $\mathbf{v}$ is the velocity of light.L Thus one cannot write : $\mathbf{p} = m\mathbf{v}$. But the equality : $\mathbf{p} = \frac{E}{c^2}\mathbf{v}$ still holds with a constant energy E, which is the case in a coulombian electric field. Hence we find the *Poincaré equation* (3.1) with a minus sign because we have chosen the left monopole :

$$\frac{d\mathbf{p}}{dt} = -\lambda \frac{1}{\mathbf{r}^3}\mathbf{p} \times \mathbf{r}\ ;\ \lambda = \frac{ecg}{E} \tag{10.8}$$

The right monopole cannot be deduced from the former by changing the sign of charge but it can be deduced by changing the sign of *the phase of the wave*, with the same magnetic charge [8].

**11. The quantum problem of a monopole in an electric central field. Angular eigenfunctions. Dirac's condition.**

We assume W = 0 and introduce the expression (5.4) of **B** in (8.4). We find the following integrals of motion [13] ( with : $D = \frac{eg}{\hbar c}$ ; $B = eB$ ) respectively for left and right monopole). The *Dirac number* $D$ was defined in (5.2) :

$$\mathbf{J}_\xi = \hbar\left[\mathbf{r} \times (-i\nabla + D\mathbf{B}) + D\frac{\mathbf{r}}{r} + \frac{1}{2}\mathbf{s}\right] ; \quad \mathbf{J}_\eta = \hbar\left[\mathbf{r} \times (-i\nabla - D\mathbf{B}) - D\frac{\mathbf{r}}{r} + \frac{1}{2}\mathbf{s}\right] \quad (11.1)$$

$\mathbf{J}_\xi$ and $\mathbf{J}_\eta$ only differ by the sign of $D$ (the sign of the eigenvalues of the charge operator). We chose the sign plus, the left monopole, and we drop the index ξ. We find:

$$[J_2, J_3] = i\hbar J_1 ; \quad [J_3, J_1] = i\hbar J_2 ; \quad [J_1, J_2] = i\hbar J_3 \quad (11.2)$$

Now, if we write **J** as :

$$\mathbf{J} = \hbar\left[\Lambda + \frac{1}{2}\mathbf{s}\right] ; \quad \Lambda = \mathbf{r} \times (-i\nabla + D\mathbf{B}) + D\frac{\mathbf{r}}{r} \quad (11.3)$$

we recognize that $\hbar\Lambda$ is *the quantum form of the Poincaré integral* (3.2). **J** is the sum of this first integral and of the spin operator : **J** is the *total angular momentum* of the monopole in an electric coulombian field, the exact analogue of the corresponding classical quantity. Of course, the components of $\hbar\Lambda$ obey the same relations (11.2) as the components of **J** .

Expressing by (5,4) **B** in terms of polar angles, from the definition (11.3) we find :

$$\Lambda^+ = \Lambda_1 + i\Lambda_2 = e^{i\varphi}\left(i\cot\theta\frac{\partial}{\partial\varphi} + \frac{\partial}{\partial\theta} + \frac{D}{\sin\theta}\right)$$

$$\Lambda^- = \Lambda_1 - i\Lambda_2 = e^{-i\varphi}\left(i\cot\theta\frac{\partial}{\partial\varphi} - \frac{\partial}{\partial\theta} + \frac{D}{\sin\theta}\right) \quad (11.4)$$

$$\Lambda_3 = -i\frac{\partial}{\partial\varphi}$$

It is important to note that, owing to our choice of gauge (5.5), there is *not any additional term* in $\Lambda_3$ as it occurred with the Dirac solution [24], [25]. Now, we look for the eigenstates $Z(\theta, \varphi)$ of $(\Lambda)^2$ and $\Lambda_3$ . In accordance with (11.2), the corresponding eigenvalue equations are :

$$(\Lambda^2)Z = j(j+1)Z ; \quad \Lambda_3 Z = mZ ;$$
$$j = 0, \frac{1}{2}, 1, \frac{3}{2}, 2, \ldots ; \quad m = -j, -j+1, \ldots, j-1, j \quad (11.5)$$

To simplify the calculation, let us introduce an angle χ and a function $D(\theta, \varphi, \chi)$ :

$$D(\theta, \varphi, \chi) = e^{iD\chi}Z(\theta, \varphi) \quad (11.6)$$

These functions are eigenstates of operators $R_k$ , as can be seen on (11.4) :

$$R^+ = R_1 + i R_2 = e^{i\varphi}\left(i\cot\theta\,\frac{\partial}{\partial\varphi} + \frac{\partial}{\partial\theta} - \frac{i}{\sin\theta}\,\frac{\partial}{\partial\chi}\right)$$

$$R^- = R_1 - i R_2 = e^{-i\varphi}\left(i\cot\theta\,\frac{\partial}{\partial\varphi} - \frac{\partial}{\partial\theta} - \frac{i}{\sin\theta}\,\frac{\partial}{\partial\chi}\right) \qquad (11.7)$$

$$R_3 = -i\,\frac{\partial}{\partial\varphi}$$

The functions $D(\theta, \varphi, \chi)$ are related by (11,6) to the eigenvalues of $\Lambda^+$, $\Lambda^-$, $\Lambda_3$ :

$$D(\theta, \varphi, \chi) = j(j+1) D(\theta, \varphi, \chi); \quad R_3 Z = m\, D(\theta, \varphi, \chi) \qquad (11.8)$$

The $R_k$ are the *infinitesimal operators of the rotation group* written in the fixed reference frame. $\theta, \varphi, \chi$ are the Euler angles : nutation, precession and proper rotation. The role of the rotation group is evident because of the *spherical symmetry* of the problem.

*Our eigenfunction problem is trivialy solved by the hypothesis of continuity of the wave functions, on the rotation group* instead of the cumbersome calculations of the so called "monopole harmonics" [24], [25], which actually don't exist ! Owing to the continuity the effects of the Dirac strings are « rubbed out » as was said at the begining.

Under the assumption of *continuity on the rotation group*, we find that the angular eigenfunctions are the *generalized spherical functions*, i.e. the matrix elements of the irreducible unitary representations of the rotation group [8], [14], [26], [27].

And they are the **eigenfunctions of the spherical top.** This was condidered by Tamm [28] as a coincidence, but here, it is evident as a consequence of the analogy between the system of a monopole in a central field and the *angular motion of a symmetrical top*.

The eigenstates of $R^2$ and $R_3$ are given by the group theory. The end of the calculation and the *radial part* may be found in [8], [13]. The most important point appears on the formula (11,6) : the $D(\theta, \varphi, \chi)$ are the elements $D_j^{m'\,m}(\theta, \varphi, \chi)$ of the unitary representations of the rotation group. So the following eigenvalues $j$, $m$, $m'$ result with :

$$j = 0, 1, \frac{1}{2}, 1, \frac{3}{2}, 2, \ldots; \quad m, m' = -j, -j+1, \ldots, j-1, j \qquad (11.9)$$

$j$ are the values of the *total angular momentum* and $m'$ is its projection on the symmetry axis of the system, joining the monopole and the coulombian center.

But $m'$ must be identical to the number $D$ in factor of $\chi$ of the exponent in (11,6). So, we have : $D = m'$ and we know from (11,1) that $D$ is the Dirac Number, thus we find:

$$D = m' = \frac{eg}{\hbar c} = -j, -j+1, \ldots, j-1, j \qquad (11,10)$$

This is the Dirac formula, but with some differences :

1) The proof is based on a model which allows an interpretation of the abstract number $n$ in (5,2).
2) The number $\hbar m'$ is limited by the quantum state of the *« top »*, which raises the question of the generality of the Dirac formula [29].

Now, the normalized angular eigenfunctions take the form [8], [14] :

$$Z_j^{m'\,m}(\theta, \varphi) = \sqrt{2j+1}\; D_j^{m'\,m}(\theta, \varphi, 0)(i)^{m'-m} \qquad (11.11)$$

The proper rotation angle $\chi$ disappears because the monopole is supposed to be *punctual*, contrary to the symmetric top. But there is a projection of the orbital momentum different from zero, due to the chirality of the magnetic charge.

**12. A nonlinear massive monopole.**

Until now we had a *massless* linear monopole (6.23), but there are *nonlinear chiral invariant* generalizations ([7], [8], [14]). We have found that the general mass is a function $F(\rho)$ where $\rho$ is given by (6.9). In Weyl's representation the Lagrangian reads:

$$L = \frac{hc}{i} \left\{ \xi^+ \left( \frac{1}{2}\frac{1}{c}[\partial_t] - \frac{g}{hc} W \right) \xi - \xi^+ \mathbf{s}. \left( \frac{1}{2}[\nabla] + \frac{g}{hc} \mathbf{B} \right) \xi \right\} + $$
$$+ \frac{hc}{i} \left\{ \eta^+ \left( \frac{1}{2}\frac{1}{c}[\partial_t] + \frac{g}{hc} W \right) \eta + \eta^+ \mathbf{s}. \left( \frac{1}{2}[\nabla] - \frac{g}{hc} W \right) \eta \right\} + hc\, F(\rho) \tag{12.1}$$

which gives the equations :

$$\frac{1}{c}\partial_t \xi - \mathbf{s}.\nabla\xi - i\frac{g}{hc}(W + \mathbf{s}.\mathbf{B})\xi + i\kappa(\rho)\sqrt{\frac{\eta^+\xi}{\xi^+\eta}}\eta = 0$$
$$\frac{1}{c}\partial_t \eta + \mathbf{s}.\nabla\eta + i\frac{g}{hc}(W - \mathbf{s}.\mathbf{B})\eta + i\kappa(\rho)\sqrt{\frac{\xi^+\eta}{\eta^+\xi}}\xi = 0 \; ; \; \left(\kappa(\rho) = \frac{dF(\rho)}{d\rho}\right) \tag{12.2}$$

These equations are chiral invariant, like the linear equation, the magnetic current (7.1) is conserved and, owing to (7.2), the equations are **PTC** *invariant* [7]. Generally, the equations (12.2) are coupled, contrary to (8.4) but this coupling is not strong. The isotropic chiral currents (9.2) are separately conserved and the coupling vanishes when : $\rho = 4|\xi^+\eta| = 0$. This happens for $\xi = 0$ or $\eta = 0$, (separated chiral components), or in the Majorana case [31], that cannot be developed here [29], [30] :

$$\xi = f(\mathbf{x},t)\, s_2\, \eta^* \;\Rightarrow\; \xi = e^{i\theta(\mathbf{x},t)}\, s_2\, \eta^* \tag{12.3}$$

Now, in (12.2), $\xi$ and $\eta$ are phase independent. The plane waves are :

$$\xi = a\, e^{i(\omega t - \mathbf{k}.\mathbf{r})} \; ; \; \eta = b\, e^{i(\omega' t - \mathbf{k}'.\mathbf{r})} \tag{12.4}$$

which gives the dispersion relation [7], [8], [14]:

$$\left(\frac{\omega^2}{c^2} - \mathbf{k}^2\right)\left(\frac{\omega'^2}{c^2} - \mathbf{k}'^2\right) - 2\left(\frac{\omega\omega'}{c^2} - \mathbf{k}\mathbf{k}'\right)\kappa^2(\rho) + \kappa^4(\rho) = 0 \; ; \; \kappa(\rho) = \frac{dF(\rho)}{d\rho} \tag{12.5}$$

We shall consider the case of an equation homogeneous in $\xi$ and $\eta$:

$$F(\rho) = \kappa_0 \rho; \; \kappa(\rho) = \kappa_0 = Const \tag{12.6}$$

Two kinds of waves (12.5) are particularly interesting :

1) $\omega = \omega'$, $\mathbf{k} = \mathbf{k}'$ : both monopoles have the same phase and the dispersion relation is reduced to :

$$\frac{\omega^2}{c^2} = k^2 + \kappa_0^2 \; ; \; \left(k = \sqrt{\mathbf{k}^2}\right) \tag{12.5}$$

This is the ordinary dispersion relation of a massive particle : a *bradyon*.

2) $\omega = -\omega'$, $\mathbf{k} = -\mathbf{k}'$. The phases have opposite signs and the dispersion relation becomes :

$$\frac{\omega^2}{c^2} = k^2 - \kappa_0^2 \tag{12.6}$$

This is the dispersion relation of a supraluminal particle, a *tachyon*. The wave equations (12.2) seem to be the first ones in which tachyons appear without any ad hoc condition. These nonlinear equations can be evaluated in various ways which in detail are described in the papers quoted in the References, especially [7].

Nevertheless, let us conclude with an important remark concerning *the nonlinear monopole in a coulombian electric field*. Chiral components of (12.2) cannot be separated as they were in the linear case (8.4).

We must go back to the $\Psi$ representation (6.23) that gives equivalently to (12.2):

$$\gamma_\mu \left( \partial_\mu - \frac{g}{\hbar c} \gamma_5 \mathbf{B}_\mu \right) \Psi + \kappa(\rho) \frac{\Omega_1 - i\gamma_5 \Omega_2}{\sqrt{\Omega_1^2 + \Omega_2^2}} = 0; \qquad \left( \rho = \sqrt{\Omega_1^2 + \Omega_2^2} \right) \tag{12.7}$$

In a coulombian electric field, with a pseudo-potential (4.7), the angular operator corresponding to (11.1), in the $\Psi$ representation, is:

$$\mathbf{J} = \hbar \left[ \mathbf{r} \times (-i\nabla + \gamma_4 D\mathbf{B}) + \gamma_4 D \frac{\mathbf{r}}{r} + \frac{1}{2} \mathbf{S} \right]; \ \mathbf{S} = \begin{pmatrix} \mathbf{s} & 0 \\ 0 & \mathbf{s} \end{pmatrix}; \ D = \frac{eg}{\hbar c}; \mathbf{B} = eB \tag{12.8}$$

To prove that **J** is an integral of the nonlinear system, *we must go back to the definition* and verify that the *mean value* of the operator **J** is a constant in virtue of the wave equations (12.7). It is just what happens and one finds indeed:

$$\frac{\partial}{\partial t} \int \Psi^+ \mathbf{J}\, \Psi \, dx\, dy\, dz = 0 \tag{12,9}$$

So, the nonlinear equation (12.7) defines the same angular momentum as the linear equation (6.23). Therefore, the angular part must be the same as in the linear case. The difference will be only in the radial factor.

### 13. Chiral gauge and twisted space.

Let us take the particular case of (12.7) when $B_\mu = O$; $\kappa(\rho) = \lambda\rho$, $\lambda = const$:

$$\gamma_\mu \partial_\mu \Psi + \lambda (\Omega_1 - i\gamma_5 \Omega_2) \Psi = 0 \tag{13.1}$$

Equivalent equations were considered by other authors ([32] - [38]), among whom was Rodichev [38] who considered a space with *affine connection*. Let us briefly recall:

1) No metric is introduced, the theory is formulated in terms of *connection coefficients* $\Gamma^i_{rk}$ only. One can define contravariant and covariant vectors $T^i$ and $T_i$, and *covariant derivatives*:

$$\nabla_\mu T^i = \partial_\mu T^i + \Gamma^i_{r\mu} T^r; \quad \nabla_\mu T_i = \partial_\mu T_i - \Gamma^r_{i\mu} T_r \tag{13.2}$$

2) Two important tensors are so defined[4], *curvature* and *torsion*:

$$-R^i_{qkl} = \frac{\partial \Gamma^i_{ql}}{\partial x^k} - \frac{\partial \Gamma^i_{qk}}{\partial x^l} + \Gamma^i_{pk}\Gamma^p_{ql} - \Gamma^i_{pl}\Gamma^p_{qk} \quad \text{and} \quad S^\lambda_{[\mu\nu]} = \Gamma^\lambda_{\mu\nu} - \Gamma^\lambda_{\nu\mu} \tag{13.3}$$

3) A *parallel transport* along a curve $x(t)$ is defined by: $\nabla_\xi T = \xi^k \nabla_k T = 0$, $(\xi = \dot{x}(t))$. A *geodesic line* is generated by the parallel transport of its tangent. Apart from an euclidian space, a *geodesic rectangle is* broken by a gap in two terms: the first, in $dt^2$, depends on torsion, the second, of the order of $o(dt^3)$, depends on curvature.

---

[4] When: $R^i_{qkl} = S^\lambda_{[\mu\nu]} = 0$, the space is euclidian.

4) In a *twisted space* ($S^\lambda_{[\mu\nu]} \neq 0$), a geodesic loop is an arc of helicoid with a « thread » of the *second order*, the order of an area. Something similar happens in a *spin fluid* : the angular momentum of a droplet is of higher order than the spin [39], [40], [41].

Now, Rodichev takes *a flat twisted space,* with torsion $\left(\Gamma^\lambda_{[\mu\nu]} = S^\lambda_{[\mu\nu]} \neq 0\right)$ but straight geodesics ($\Gamma^\lambda_{(\mu\nu)} = 0$), and the following connection and covariant spinor derivative :

$$\Gamma_{\lambda[\mu\nu]} = S_{\lambda\mu\nu} = \Phi_{[\lambda\mu\nu]} \; ; \; \nabla_\mu \Psi = \partial_\mu \Psi - \frac{i}{4} \Phi_{[\mu\nu\lambda]} \gamma_\nu \gamma_\lambda \Psi \tag{13.4}$$

with the following Lagrangian density :

$$L = \frac{1}{2} \left\{ \bar\Psi \gamma_\mu \nabla_\mu \Psi - \left(\nabla_\mu \bar\Psi\right) \gamma_\mu \Psi \right\} \tag{13.5}$$

Translating the last formula in our language, it gives:

$$L = \frac{1}{2} \left\{ \bar\Psi \gamma_\mu \partial_\mu \Psi - \left(\partial_\mu \bar\Psi\right) \gamma_\mu \Psi - \frac{i}{2} \Phi_{[\mu\nu\lambda]} \bar\Psi \gamma_\nu \gamma_\lambda \Psi \right\} \tag{13.6}$$

Introducing the axial dual vector : $\Phi_\mu = \frac{i}{3!} \varepsilon_{[\mu\nu\lambda\sigma]} \Phi_{[\nu\lambda\sigma]}$, the lagrangian becomes :

$$L = \frac{1}{2} \left\{ \bar\Psi \gamma_\mu \partial_\mu \Psi - \left(\partial_\mu \bar\Psi\right) \gamma_\mu \Psi - \frac{1}{2} \Phi_\mu \bar\Psi \gamma_\mu \gamma_5 \Psi \right\}, \tag{13.7}$$

which gives the equation :

$$\gamma_\mu \left( \partial_\mu - \frac{1}{2} \Phi_\mu \gamma_5 \right) \Psi = 0 \tag{13.8}$$

With $\Phi_\mu = \frac{2g}{\hbar c} B_\mu$, this is *our equation* (6.23). Let us note that Rodichev did not introduce $\Phi_\mu$ as an external field, but only as an geometric property, but now *we can say that a monopole plunged into an electromagnetic field induces a torsion in the surrounding space*.

Rodichev ignored the monopole. He didn't aimed at the linear equation (13.6), but at a nonlinear equation, through the following Einstein-like action integral without external field :

$$S = \int (L - bR) d^4x, \tag{13.9}$$

$L$ is given by (13.3), $b = Const$, $R = $ total curvature and, in virtue of (13.3) :

$$R = \Phi_{[\lambda\mu\nu]} \Phi_{[\lambda\mu\nu]} = -6 \Phi_\mu \Phi_\mu \tag{13.10}$$

Hence, (13.9) becomes :

$$S = \int \left\{ \frac{1}{2} \left[ \bar\Psi \gamma_\mu \partial_\mu \Psi - \left(\partial_\mu \bar\Psi\right) \gamma_\mu \Psi - 2\Phi_\mu \bar\Psi \gamma_\mu \gamma_5 \Psi \right] + 36b \; \Phi_\mu \Phi_\mu \right\} d^4x \tag{13.11}$$

Now, if we vary $S$ with respect to $\Phi$, we find :

$$\Phi_\mu = \frac{1}{18b} \bar{\Psi}\gamma_\mu\gamma_5\Psi$$
$$R = \frac{1}{64\beta^2} \left(\bar{\Psi}\gamma_\lambda\gamma_\mu\gamma_\nu\Psi\right)\left(\bar{\Psi}\gamma_\lambda\gamma_\mu\gamma_\nu\Psi\right) = \frac{3}{32\beta^2} \left(\bar{\Psi}\gamma_\mu\gamma_5\Psi\right)\left(\bar{\Psi}\gamma_\mu\gamma_5\Psi\right) \qquad (13.12)$$

Now, the variation of $S$ with respect to $\Psi$ gives the non linear equation :

$$\gamma_\lambda\partial_\lambda\Psi - \frac{1}{9b^2}\left(\bar{\Psi}\gamma_\mu\gamma_5\Psi\right)\gamma_\mu\gamma_5\Psi = 0 \qquad (13.13)$$

So doing, *we come back once more to the monopole,* but now *in the nonlinear case.* Up to a constant factor, (13.13) is identic to (13.1), a particular case of (12.7). The identity between (13.3) and (13.1) is due to the identities (6.11), in virtue of which, and of (13.12) :

$$R = \frac{3}{32\beta^2}\left(\Omega_2^2 + \Omega_2^2\right) \qquad (13.14)$$

Which means that **the fundamental chiral invariant, $\left(\Omega_2^2 + \Omega_2^2\right)$, apart from a constant factor, is the curvature of the twisted space created by the self action of the monopole**, expressed in the equation by the identification of the torsion to the total curvature in (13.8). Which confirms the link between our monopole and a torsion of the space.

**14. The electroweak generalisation by Stumpf.**

We owe to H. Stumpf an important generalization of the preceding theory, which could not be better summarized than by quoting the formulation of the problems by the author himself:

*(i) Does a medium exist which transmits electric as well as magnetic monopole actions ?*
*(ii) Can one discover « elementary » or other particles which act as magnetic monopoles or dyons, respectively ?*
*(iii) Can the hypothetical medium and the monopoles and dyons be incorporated into an extended electroweak Standard model ?* [49].

In this context it is interesting to note that in de Broglie's theory of fusion the problem of the existence of magnetic monopoles is already present in the formalism. Apart from « electric » photons the fusion equations admit a second photon solution which has been identified as a « magnetic » photon state, [8], [43][5], the fields of which are exactly those that enter in the dynamics of a magnetic charge. In this way the problem of magnetic monopoles is linked to the fermionic substructure of the photon, or more general, to the substructure of elementary particles. This has been the topic of de Broglie's and Heisenberg's fusion ideas. Following these ideas Stumpf developed a quantum field theoretic formalism for the treatment of fusion problems and in particular, he applied this formalism to the monopole problem. A discussion of his results would exceed the scope of the paper. So we refer to the literature [42]-[50].

**15. Experiments.**

Most experiments were performed in Moscow in the Recom Laboratory of the Kurchatov Institute, under the leadership of Leonid Urutskoiev [51], [53[1]], some at the Nuclear Institute of Dubna, by Vladimir Kuznetsov et alter [52], and others at the Kazan University, by Nikolai Ivoilov [53[2]], [55]. At first we describe Urutskoiev's experiments, performed with intense, brief electrical discharges through thin titanium electrodes submerged in a liquid medium (generally water). He found several remarkable effects :

1) The appearance of an astonishingly stable *lightning ball* (50 times the duration of the discharge) with a very complex optical spectrum, showing the rays of various chemical elements, many of which were initially absent from the laboratory installation [51], [53[1]].

---

[5] Two references can be added, concerning « electric » and « magnetic » photons :
G ; Lochak, Ann. Fond. L. de Broglie, **20**, 1995, p. 111 ; **29**, 2004, p. 297.

2) The most remarkable effect was the *chemical composition* of the remaining dust of the thin titanium electrode pulverized by the electric discharge: the complex composition obtained by *mass spectrography* confirmed the one obtained by optical spectrometry : Fig. 3 (the chemical components present before the experiment are not shown).

An important point is a *modification of the isotopic spectrum* of the elements *: the proportions of the different isotopes are modified.* It is the case for the titanium, the central isotope of which is strongly weakened, but it must be stressed (Urutskoiev) that it is not transformed into the other isotopes. One can see on the Fig. 4 that the central $^{48}Ti$ isotope is strongly weakened, indeed, but the lateral satellites are practically unaltered,

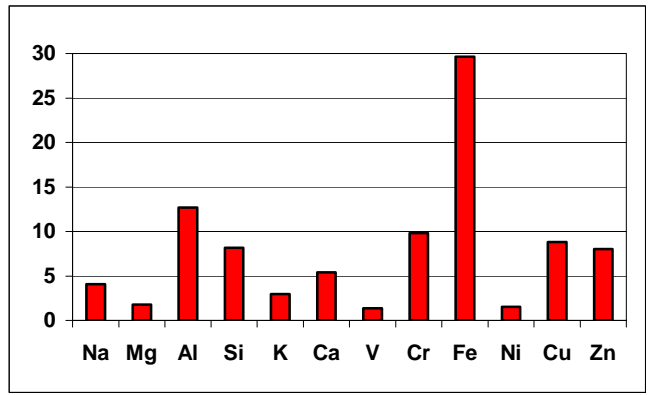

Fig. 3. New elements after the discharge (Urutskoiev)

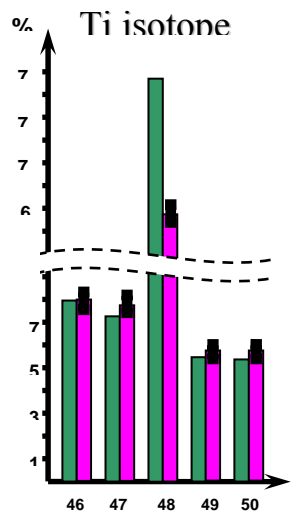

Fig. 4. Isotopic structure of titanium before and after the discharge (Urutskoiev)

An interesting fact is the presence of a considerable quantity of iron (Fig. 3) that could pass for an artefact, since the source of monopoles was in a block of steel. So that this point was specially verified. The isotopic composition of the column "iron" that appears in Fig. 3 *is not the ordinary composition of iron.* For instance, the iron $^{56}Fe$ abundant in nature was strongly reduced, while the iron $^{57}Fe$, very rare in nature (less than 2,5 %) was strongly increased. Thus, this iron is not the one that enter in the composition of the source.

It must be stressed that exceptionally competent and scrupulous experimenters obtained these results. The Urutskoiev group repeated and controlled hundreds of similar discharge experiments, which were in addition confirmed on several mass spectrometers of different types in different laboratories, mainly by the Kuznetsov group of Dubna [52].

3) A puzzling result was that the radiation emitted during the electrical discharge was examined on nuclear photographic plates located at distances of several meters from the source. **S**trange tracks appeared on the plates**.** Figures like Fig. 5, communicated by Urutskoiev, were analyzed by specialists skilled in interpreting tracks on nuclear emulsions.

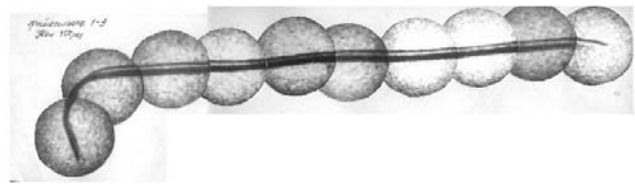

Fig. 5. A monopole track (Urutskoiev)

The conclusion was that these tracks were unlike anything they had ever observed — don't forget that the Kurchatov Institute is one of Russia's major nuclear physics laboratories !

These tracks could not be due to electrically charged particles, because:

a) The observed particles freely cross several meters of atmosphere (it was not done in vacuum), while electric particles would be largly stoped.

b) For electric particles, the track thickness would correspond to a 1 Gev energy, but the tracks were 'hairless': without surrounding **"**delta electrons", characteristic of charged particles.

c) They cannot be neutral since they leave tracks. Thus, they must carry some other charge.

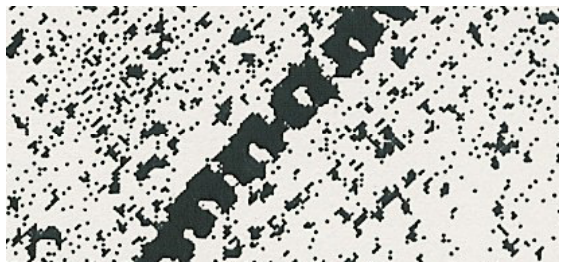

Fig. 6. Caterpillar structure of a monopole track (enlarged $\cong$ 150 times).

These tracks have a curious 'caterpillar structure' (Fig. 6 and references quoted above) The formation of the tracks is sensitive to a magnetic field. A field of 20 œrsteds applied to the source of monopole radiation transforms the shape of the tracks into a broader trace of 'comet-like' shape *with an integrated darkening equal to that of the initial track.*

Another question is raised by another specific feature : the traces appear in a plane orthogonal to the radius-vector from the center of the unit, as if they were traped between the film and the sensitive emulsion. The larger is the distance between the detector and the unit center, the narrower is the trace pattern. At a distance equal to about half meter, the track width is about 30 μm, while at a 2-meter distance it is only around 5 μm." [53$_1$]

4) Various difficulties of interpretation gradually led Urutskoiev and his research team to the conclusion that magnetic poles could be a possible source of the strange radiation effects they had observed. They became aware of the present author's work and a fruitful collaboration has been initiated.

From the very beginning, an important experiment was realized by Urutskoiev and Ivoilov [54], using the fact that $^{57}Fe$ is at the same time *magnetic* and *the most sensitive element to the Mössbauer effect*. They irradiated, at some meters from the source of the supposed monopoles, a sample of $^{57}Fe$. Behind the iron sample was one pole of a long linear magnet, in order to repel the monopoles of the same sign and attract the monopoles of the opposite sign. Owing to the Mössbauer effect, *they found a distinct shift of a characteristic $\gamma$ ray*.

They repeated the experiment with the *other pole* of the magnet behind the iron sample and, with the same exposure they found a $\gamma$ ray shift in the opposite direction [54].

One can make two remarks about this experiment :

a) This is one of the most brilliant proof of monopole magnetism. But there are others : for instance, the fact that Ivoilov focused a monopole beam with an electromagnet.

b) If the $^{57}Fe$ target sample used in the Mössbauer experiment is abandoned for three days, the preceding characteristic $\gamma$ ray spectrum goes back to its mean normal position. This half-life effect seems to hold for all the effects of magnetism induced by monopoles: they seem to have a limited time of life (not predicted by theory). But other effects, such isotopic shifts are definitive.

**More recent experiments:**

1) **Chemical effects:** Urutskoiev has decomposed a sample of *ammonium nitrate* $(NH_4NO_3)$ sealed in a hermetic aluminum vessel at a distance of several meters from the electric discharge, hypothesizing that the monopoles emitted by the discharge would penetrate the aluminum container and catalyze the exothermic decomposition. He further introduced the same material in a vessel made of ferromagnetic steel, and nothing happened. The experiment was repeated many times for statistical accuracy. The nitrate was decomposed in every aluminum container and no change was ever observed inside steel containers.

2) **Enrichment of Uranium:** The first idea of a catastrophe caused by a flow of monopoles arose in the Urutskoiev group after the Chernobyl catastrophe. The hypothesis about Chernobyl is that the origin could be an electric "machine explosion" that really happened in a building connected to the reactor by a water pipe. Urutskoiev hypothesized that monopoles created by the arc could have been conveyed into the reactor through this water conduit and perpetrated transmutations on a massive scale.

The original concentration of $^{235}$U, before the catastrophe, in 1986, was less than 1.2%. But after the explosion, uranium pieces were found, enriched up to 27%. Later, Urutskoiev tried to obtain an identical effect on uranium with a monopole flow. He proved that it is a repeatable effect, and an impressive enrichment was obtained in his laboratory [53$_1$].

3) $\beta$ **Radioactivity**. There are two important results:

a) Il was theoretically predicted [8], [12], [14], that in some cases, $\beta$ radioactivity may be associated with monopoles, instead of neutrinos, and that solar monopoles could follow the earth magnetic lines and reach the magnetic poles. Our initial hypothesis was confirmed by experiments performed by Ivoilov [53$_2$], [55]. He submitted a $\beta$ radioactive sample to a magnetic field in the presence of nuclear emulsions and monopole traces appeared.

b) *Reduction of the half-life time of $\beta$ emitters by magnetic monopoles.* It is known that the lifetime of unstable nuclear states of some atoms depend of their chemical state ([56], [57] [58]). While this effect is not controversial, the changes are generally less than 1% and are close to the noise level. However, the $\beta$ half-life of rhenium has been reduced from $4 \times 10^{10}$ years to 30 years in the case of a complete ionization of the atom.

Urutskoiev put this phenomenon in parallel with the Kadomtsev effect, which consists in a reduction of a $\beta$ lifetime by the action of an external magnetic field on an atom. This theoretical prediction [59] needs gigantic fields (unobtainably large on a laboratory scale). But such a field can be produced in the vicinity of a monopole, which is able to pass very close to a nucleus because there is no repelling force. Experiments performed by Urutskoiev confirm this interaction.

4) **Chirality**. Ivoilov, with a weaker source of monopoles, has the curious advantage to obtain the same tracks with lower energy, more sensitive to the distortions of fields, thus with more complicated and easily recognizable shapes. The Fig. 7 shows a monopole track with its image in a mirror made of glass or monocrystalline Si or Ge [53$_2$], [55].

Fig. 7. A monopole track with its image in a mirror (Ivoilov)

The photograph is taken on a low sensitivity *two-sided* X-ray plate, so that a microscope of magnification 20 to 100 is able to distinguish the track on the side of the source of radiation from the track on the reflector side. They are almost identical, *up to little defects* due to the fluctuating magnetic field due to the electric arc of the source of monopoles, from which the sensitive emulsion plate was at a distance of only 10-15cm.

There are three questions concerning this experiment :

a) As we already know, the observed tracks are on a plane *orthogonal* to the direction of the source. Perhaps Ivoilov provided the beginning of an answer to this puzzling fact (partially in [53$_2$], partially in a private discussion). He quoted that there is a too little number of tracks, to explain macroscopic effects such as the magnetization of a sample of $^{57}Fe$ or the fact that fragments of irradiated titanium are attracted by a magnet after the electrical discharges.

Conversely, Ivoilov noted on the photographic plates myriads of *microscopic tracks* that did not draw attention at a first glance, as they looked as small defects of the surface. On closer inspection, however, it turned out that the microscopic tracks were not simple defects. Ivoilov verified that these traces were produced by the electric discharge, and he emitted the hypothesis that perhaps they are due to monopoles passing through the emulsion plane at greater angles. The long tracks, which can reach 10 mm, would then be the paths of rare monopoles that are trapped by chance *between the sensitive layer and the polymer ground* (the last point seems to be verified). And he stressed that this faculty of making a bend at a right angle could be a consequence of the *zero mass* predicted by theory [53$_2$], [55] .

2) *On the Fig.7 the reflected track is not a mirror image in the optical sense : it is quasi-identical to the direct track*, which has been verified on many photographs. This is a first proof that the monopole is a pseudo scalar, as it was predicted by Pierre Curie and fonfirmed by the theory given above. It must be stressed that this theory is the only one, which deduces the chirality of monopoles from its wave equations (8.4), splitting the south and north monopoles. Other theories, devoid of geometry, cannot integrate this result.

3) Last question: *the reflected track seems to be rotated of an angle $\pi$*, which is puzzling at a first glance, but actually the image is not rotated, it *is symmetric with respect to a center.* Contrary to the rotation, this symmetry is evident and achieves the proof of pseudo-scalarity of the charge : the magnetic current is a pseudo-vector and has the symmetry of the product of the velocity (a polar vector) by the charge (a pseudo-scalar). The component orthogonal to the plane of symmetry does not change, while the components parallel to the plane change their sign. So that the experimental result agrees with the theoretical predictions

4) **Biological effects.** These researches were organized under the leadership of E.A. Pryakhin (from the *Ural Scientific and Practical Center of Radioionic Medicine*), in the Urutskoiev Laboratory, where an experimental sample of several hundred laboratory mice were irradiated by magnetic monopoles. The conclusions of Pryakhin's report at a conference in Marseille in 2004 [$53_3$], draw attention to the following points:

" It can be concluded, based on the results of our experiments, that:

*1. "Strange" radiation stimulates proliferation of bone marrow cells with or without delay in maturation;*

*2. It induces changes resulting in increased resistance to genotoxic exposures (gamma-irradiation and others);*

*3. "Strange" radiation" (monopoles) aggravates the clinical course of acute radiation disease if it is applied after gamma-irradiation.*

*4. It leads to changes of cell composition in the blood* [$53_3$]".